# Orientational Order Governs Collectivity of Folded Proteins*

Canan Atilgan, Osman Burak Okan, Ali Rana Atilgan

Faculty of Engineering and Natural Sciences, Sabanci University, 34956 Istanbul Turkey

**ABSTRACT**

The past decade has witnessed the development and success of coarse-grained network models of proteins for predicting many equilibrium properties related to collective modes of motion. Curiously, the results are usually robust towards the different cutoff distances used for constructing the residue networks from the knowledge of the experimental coordinates. In this study, we present a systematical study of network construction, and their effect on the predicted properties. Probing bond orientational order around each residue, we propose a natural partitioning of the interactions into an essential and a residual set. In this picture the robustness originates from the way with which new contacts are added so that an unusual local orientational order builds up. These residual interactions have a vanishingly small effect on the force vectors on each residue. The stability of the overall force balance then translates into the Hessian as small shifts in the slow modes of motion and an invariance of the corresponding eigenvectors. We introduce a rescaled version of the Hessian matrix and point out a link between the matrix Frobenius norm based on spectral stability arguments. A recipe for the optimal choice of partitioning the interactions into essential and residual components is prescribed. Implications for the study of biologically relevant properties of proteins are discussed with specific examples.

**AUTHOR SUMMARY**

Network models of proteins have opened up previously unexplored areas of study, since the level of coarse graining adopted has been shown to reveal several important characteristic of these systems. The findings are mainly based on the observation that a simplified, harmonic potential is capable of describing the collective modes of motion, which are also associated with the basic functioning of these molecular machines. The level of success of these studies depends on the quality with which we describe the interactions between pairs of residues in the protein. Here, we perform for the first time, a systematical study on how the predictions are affected by network construction. We first track the local orientational order of residues as new contacts are added with increasing cut-off distance. We also study in detail the spectral properties of the Hessian. We show why the network construction is free of the cut-off distance problem beyond a threshold value, if one is interested in the collective motions. We discuss the implications on the limitations and capabilities of the network models with regard to functionality-related predictions based on the most global motions.

---

*All computer programs used in the analyses are available upon request.

INTRODUCTION

Globular proteins show diversified structures and sizes, yet, it has been claimed that they display a nearly random packing of amino acids with strong local symmetry on the one hand [1], and that they are regular structures that occupy specific lattice sites, on the other [2]. It was later shown that this classification depends on the property one investigates, and that proteins display "small-world" properties, where highly ordered structures are altered with few additional links [3]. Furthermore, packing density of proteins scales uniformly with their size [4,5] which causes them to show similar vibrational spectral characteristics to those of solids [6]. Dynamical studies of folded proteins draw much attention to their importance in relating the structure of the proteins to their specific function and collective behavior. Protein dynamics is generally both anisotropic and collective. Internal motional anisotropy is a consequence of the low symmetry local atomic environment, while the collectivity is mainly caused by the dense packing of proteins [7].

Theoretical studies on fluctuations and collective motions of proteins are based on either molecular dynamics (MD) simulations or normal mode analysis (NMA). Since, in molecular simulations with conventional atomic models and potentials, computational effort is demanding for proteins with more than a few hundreds of residues, coarse grained models with simplified governing potentials have been employed. The latter have shown a great success in the description of the residue fluctuations and the collective behavior of proteins [8].

One of these simplified models, NMA using a single parameter harmonic potential [9] following the uniform harmonic potential introduced originally by Tirion [10], successfully predicts the large amplitude motions of proteins in the native state [11]. Within the framework of this model, proteins are modeled as elastic networks whose nodes are residues linked by inter-residue potentials that stabilize the folded conformation. The residues are assumed to undergo Gaussian-distributed fluctuations about their native positions. The springs connecting each node to all other neighboring nodes are of equal strength, and only the atom pairs within a cut-off distance are considered without making a distinction between different types of residues. This model, with its simplicity, speed of calculation and relying mostly on geometry and mass distribution of the protein, demonstrates that a single-parameter model can reproduce complex vibrational properties of macromolecular systems. Elastic models based on the force balance around each node [12] led to the development of the so called Anisotropic Network Model (ANM) [13]. In the past few years, variant methods have been introduced; e.g. [14,15]. Applications of these models on many proteins show successful results in terms of predicting the collective behavior of proteins. By separating different components of normal modes, e.g. collective (low-frequency) motions, the nature of a conformational change, for example due to the binding of a ligand, may also be analyzed [16,17].

Despite numerous applications comparing the theoretical and experimental findings on a case-by-case basis, e.g. [18,19,20], only a few attempted a statistical assessment of the models. A methodology that evaluates the number of modes necessary to map a given conformational change from the degree of accuracy obtained by the inclusion of a given number of modes, showed the results to be protein dependent [21]. In another study where 170 pairs of structures were systematically analysed, it was shown that the success of coarse-grained elastic network models may be improved by recognizing the rigidity of some residue clusters [22,23].

To date, the structures that form the basis of the network models have been generated from certain rules of thumb. A distance threshold between the $C_\alpha$ or $C_\beta$ atoms of the residues is used as the rule for the connectedness of a given pair of residues. Values in the range of 8 – 17 Å are found in the literature based on the argument that (i) the eigenvalue distributions obtained from the modal decomposition are similar to those obtained from the full-atom NMA description of proteins, or (ii) these provide atomic fluctuation profiles that display the largest correlation with the experimental B-factors.

Recently, distance weighted interaction schemes of various functional forms were proposed to circumvent the cut-off problem [23]. Voronoi tessalation of the space defined by the central (usually $C_\alpha$ or $C_\beta$) atom into non-intersecting polyhedra constitute another route that frees one from defining a cut-off distance [24]. Atom-based network construction approaches have also been used [see reference [25] for a review of the variety of network construction methods in literature.]

In this study, we use a systematic approach on a large set of globular proteins with varying architectures and sizes to find a basis for why the network models work well to define certain properties of the system. This enables us to assess residue-based approaches used in the construction of the networks. We track the local orientational order of residues as new contacts are added with increasing cut-off distance. We show that the network construction is free of the cut-off distance problem once a certain baseline threshold is accessed, if one is interested in the collective motions and the fluctuation patterns of the residues. Implications for the limitations and capabilities of the ANM methodology are discussed due to functionality-related predictions based on the most global motions.

**COMPUTATIONAL DETAILS**

**Network construction.** A protein of $N$ residues is treated as a residue-based structure, where the $C_\alpha$ atom of each amino acid is considered as a node, and the coordinates of the protein are obtained from the protein data bank (PDB) [26]. The network information is contained in the $N \times N$ adjacency matrix, **A**, of inter-residue contacts, whose elements $A_{ij}$ are taken to be 1 for contacting pairs of nodes $i$ and $j$, and zero otherwise. The criterion for contact is that the two nodes are within a cut-off distance $r_c$ of each other.

**Bond Orientational Order Parameter**. We use the bond-orientational order parameter defined by Steinhard et al., a well established metric in the study of packed spheres [27], to analyze local connectivity around each residue:

$$Q_l(i) \triangleq \left( \frac{4\pi}{2l+1} \sum_{m=-l}^{l} \left| \frac{1}{N_b(i)} \sum_{n=1}^{N_b(i)} Y_{lm}(\theta(\vec{r}_n - \vec{r}_i), \phi(\vec{r}_n - \vec{r}_i)) \right| \right)^{1/2} \quad (1)$$

$Y_{lm}(\theta(\vec{r}_n - \vec{r}_i), \phi(\vec{r}_n - \vec{r}_i))$ are the spherical harmonic functions for a bond vector from residue $i$ to $n$, $\theta$ and $\varphi$ are the polar angles of this bond. $N_b(i)$ is the total number of such contacts of $i$. This form of the order parameter is invariant under reorientations of the external coordinate system. In defining a mean protein environment, we further average $Q_l(i)$ over all residues;

$$\langle Q_l \rangle = \frac{1}{N} \sum_{i=1}^{N} Q_l(i) \quad (2)$$

Among different choices for $l$, $Q_{l=6}$ is commonly employed as the bond orientational parameter, since it concurrently yields non-zero values for hexagonal close packed, cubic (simple, body centered and face centered) and icosahedral configurations [27,28,29].

**Anisotropic network model.** In ANM**,** the networks are formed as described under the subsection *Network Construction* and the interactions between nodes is considered to be due to harmonic potentials [13]. Nodes within the predetermined cut-off distance $r_c$ are coupled by elastic springs having a uniform force constant $\gamma$. Thus, the overall potential of the molecule is given by the sum of all harmonic potentials among interacting nodes such that

$$V = \frac{\gamma}{2} \sum_i \sum_{j>i} A_{ij}(R_{ij} - R_{ij}^0)^2 \qquad (3)$$

$R_{ij}^0$ is the average distance between residues $i$ and $j$. For a network of $N$ nodes, the Hessian is a $3N \times 3N$ matrix formed by a number of $N^2$ super elements $\mathbf{H}_{ij}$. The off-diagonal super elements of $\mathbf{H}_{ij}$ ($i \neq j$), obtained from the second derivative of the total potential with respect to node positions, are given by

$$\mathbf{H}_{ij} = \frac{\gamma A_{ij}}{(R_{ij}^0)^2} \begin{bmatrix} X_{ij}X_{ij} & X_{ij}Y_{ij} & X_{ij}Z_{ij} \\ Y_{ij}X_{ij} & Y_{ij}Y_{ij} & Y_{ij}Z_{ij} \\ Z_{ij}X_{ij} & Z_{ij}Y_{ij} & Z_{ij}Z_{ij} \end{bmatrix} \qquad (4)$$

where $X_{ij}$, $Y_{ij}$, and $Z_{ij}$ are the Cartesian components of the distance vector $R_{ij}^0$. The diagonal super elements are given by $\mathbf{H}_{ii} = -\sum_{j, j \neq i} \mathbf{H}_{ij}$.

The elements of the inverse of the Hessian, $\mathbf{G} = \mathbf{H}^{-1}$, may be used to predict the auto- and cross-correlations of residues. $\mathbf{G}$ may be viewed as an $N \times N$ matrix whose $ij$th element is the 3×3 matrix of correlations between the $x$-, $y$-, and $z$-components of the fluctuations $\Delta \mathbf{R}_i$ and $\Delta \mathbf{R}_j$ of residues $i$ and $j$; i.e.,

$$\mathbf{G}^{ij} = \begin{bmatrix} \langle \Delta X_i \Delta X_j \rangle & \langle \Delta X_i \Delta Y_j \rangle & \langle \Delta X_i \Delta Z_j \rangle \\ \langle \Delta Y_i \Delta X_j \rangle & \langle \Delta Y_i \Delta Y_j \rangle & \langle \Delta Y_i \Delta Z_j \rangle \\ \langle \Delta Z_i \Delta X_j \rangle & \langle \Delta Z_i \Delta Y_j \rangle & \langle \Delta Z_i \Delta Z_j \rangle \end{bmatrix} \qquad (5)$$

The correlations between residue pairs are obtained from the trace of its components, $\langle \Delta \mathbf{R}_i \cdot \Delta \mathbf{R}_j \rangle = \text{trace}(\mathbf{G}^{ij})$. In particular, the auto-correlations are the average residue fluctuations in space, and are directly proportional to the experimentally measured B-factors:

$$\langle \Delta \mathbf{R}_i \cdot \Delta \mathbf{R}_i \rangle = \text{trace}(\mathbf{G}^{ii}) \qquad (6)$$

**RESULTS**

**Structural heterogeneity of amino acid distributions in proteins.** The local environment of a residue is imposed by the spatial organization of the other residues. Studies exploring possible correspondence between various forms of local order and the amino acid packing have been long explored [1,2,30,31,32]. Obviously, there is no single matching structure; instead, one finds traces of various well-known ordered states like the icosahedral and face centered cubic arrangements. The basic premise of the ANM is the local orientational heterogeneity which establishes unique force balance constraints around each node [13]. In the cut-off based models, local anisotropy is a natural function of the coordination radius. For a better understanding of this spatial dependence, we use the bond orientational parameter $Q_6$ (equation 2). In a fully extended chain conformer, this parameter is exactly one and for a fully random distribution of bonds, it is zero. $Q_6$ has also been reported for common regular packing arrangements; e.g. it is 0.57, 0.51 and 0.35 for 13-atom Face Centered Cubic (FCC), 15-atom Body Centered Cubic (BCC) and 7-atom Simple Cubic (SC) cluster [27], respectively, within their first coordination shell.

In figure 1, we present $Q_6$ for three different proteins, as well as the average for 8940 residues pertaining to 30 different proteins. As expected, at lower cut-off distances, we find lower symmetry (higher anisotropy) in the average residue environment. At smaller cut-off choices, a better correspondence is established between the environment of an average residue and that of common cubic crystallographic arrangements (FCC, BCC, SC). For example, assuming packing of spherical particles with a diameter of 3.7 Å, equal to average nearest neighbor $C_\alpha$-$C_\alpha$ atom distances, $Q_6$ is 0.27 for FCC and SC lattices at 6 Å, while it is 0.33 for the average residues. With the addition of bonds in a larger coordination space, local anisotropy gradually decreases. This intuition is strongly validated for crystallographic configurations when further coordination shells are taken into account, where $Q_6 \approx 0$ at distances larger than 9 Å. However, for proteins, the newly formed orientational order is nontrivial and noticeably persistent above $r_c$=10 Å converging to around 0.13. The plateau which is reached by $Q_6$ hints a non-negligible residual anisotropy and is comparable in value to what has been computed for a supercooled Lennard-Jones system [27].

The non-vanishing and uniform value of the bond orientational order at larger cut-off settings suggests that the network structure is well-preserved even if we introduce new contacts to the system. The smallest distance at which this particular bond orientational order builds up can be seen as a core structure which contains the essence of structural dynamics exhibited by the molecule. Motivated by this observation, we shall partition the Hessian matrix into two, as the essential and residual parts. This simplifies the picture and brings in valuable insight for the cut-off problem as we show next.

**Partitioning the Hessian into its essential and residual components.** The distributions studied in the previous subsection lead us to further examine the effect of the geometrical features brought into the system for the different choices of contacts. The system properties of interest studied by network models all rely on the construction of the Hessian matrix. Contacts within a chosen cut-off distance are all assumed to interact identically (i.e. uniform $\gamma$ as in equation 4). Therefore, we study how the eigenvalue and eigenvector structure of the Hessian is modified by these choices.

We first partition the Hessian into two parts:

$$\mathbf{H} = \mathbf{H}_* + \mathbf{H}_r \qquad (8)$$

We postulate that $\mathbf{H}_*$ contains information due to the essential contacts of the nodes, whereas $\mathbf{H}_r$ is the residual part where the interactions are added in a spherically symmetrical manner around the nodes. We assume that there is a total of $M = \sum_i m_i /2$ interactions between pairs of nodes and $r$ of these are of the latter type. Here $m_i$ is the number of interactions of the $i^{th}$ node.

Equation 8 may be expanded as

$$\mathbf{H} = \mathbf{H}_*(\mathbf{I} + \mathbf{H}_*^{-1}\mathbf{H}_r) = \mathbf{H}_*\left(\mathbf{I} + \mathbf{U}_*\mathbf{\Lambda}_*^{-1}\mathbf{U}_*^T\mathbf{U}_r\mathbf{\Lambda}_r\,\mathbf{U}_r^T\right) \qquad (9)$$

where the eigenvalue decomposition $\mathbf{H}_* = \mathbf{U}_*\mathbf{\Lambda}_*\,\mathbf{U}_*^T$ and $\mathbf{H}_r = \mathbf{U}_r\mathbf{\Lambda}_r\,\mathbf{U}_r^T$ are used in the latter term. Inasmuch as the eigenvectors of the $\mathbf{H}_*$ and $\mathbf{H}_r$ matrices are uncorrelated, the product $\mathbf{U}_*^T\mathbf{U}_r$ and therefore $\mathbf{H}_*^{-1}\mathbf{H}_r$ will vanish and the Hessian will be $\mathbf{H} = \mathbf{H}_*$. In practice, a complete independence of the two matrices is not expected, mainly due to the finite size, and hence the surface effects, of the protein systems studied. Then these interactions are not perfectly uncorrelated, and the Hessian is approximated by $\mathbf{H} \approx \mathbf{H}_*$. In the Appendix, we outline the conditions under which this approximation may be achieved. The changes that are brought about by the $\mathbf{H}_*^{-1}\mathbf{H}_r$ term will have the result of modifying the eigenvalues of the $\mathbf{H}_*$ matrix so that their order might be affected, and the corresponding eigenvectors will be perturbed. The eigenvectors of the lowest eigenvalues will be the least sensitive to these perturbative effects as we outline below [also see Chapter 8 in [33]].

First, we map all the eigenvalues of **H** into the interval [0,1], to alleviate the size effects while comparing the spectral properties obtained with different cut-offs, and for proteins of different sizes. This is achieved by rescaling the 3 × 3 super-elements $\mathbf{H}_{ij}$ as,

$$\mathbf{H}_{ij}^{\dagger} = \frac{\mathbf{H}_{ij}}{\sqrt{\text{trace}(\mathbf{H}_{ii})}\sqrt{\text{trace}(\mathbf{H}_{jj})}} \tag{10}$$

The spectral stability theory of non-singular matrices provides upper bounds for the perturbation of eigenvectors under matrix perturbations. Let $\mathbf{p}_i$ be an eigenvector and $\mathbf{q}_i$ its perturbed counterpart. An upper bound for the angle between the two vectors is [34]

$$\frac{1}{2}\sin 2\theta_i = \frac{\|\mathbf{H}_*\|}{\text{gap}(i, \mathbf{H}_* + \mathbf{H}_r)} \tag{11}$$

The gap function gap($i$,$\mathbf{A}$) for a matrix $\mathbf{A}$ is defined as the smallest difference between a given eigenvalue and the remaining elements of the spectrum; $\min_{ij}|\lambda_i - \lambda_j|$. $\|\mathbf{A}\| = [trace(\mathbf{A}^T\mathbf{A})]^{1/2}$ is the Frobenius norm. Due to the natural singularity of the Hessian owing to rigid body motions, equation 11 is not directly applicable to compute an upper bound for $\theta$. Moreover, the value of the gap function for a given mode number $i$ depends on the cut-off distance at which the $\mathbf{H}_r$ matrix is formed. Nevertheless, the analysis pointed out by equation 11 is insightful in that the shift in a given eigenvector is expected to get lower when the Frobenius norm is minimized.

We probe the dependence of the Frobenius norm of $\mathbf{H}^{\dagger}$ on $r_c$ as shown in Figure 2 for six proteins of various sizes. For each case, we note minima by the arrow. Typically, the curves decrease from a high value at low $r_c$ to a minimum, $r_c^*$. During this initial decrease phase, the additional contacts that are included in the calculation of $\mathbf{H}^{\dagger}$ bring in non-trivial structural information that contributes to the lowering of the Frobenius norm. At $r_c$ values higher than $r_c^*$, on the other hand, the new contacts do not modify the overall orientational structure $Q_6$, hence the force balance around the nodes, of the system (figure 1). However, the monotonic addition of these uniform structural elements increases their overall weight in the Frobenius norm, resulting in an increase in the calculated values.

**Invariant eigenvectors.** To quantify the effect of the perturbation introduced by $\mathbf{H}_r$ on the eigenvectors of $H_*$, we first compute the eigenvector, $\mathbf{p}_i(r_c)$, that belongs to a selected mode $i$ at a series values spanning $r_c = 8 - 23$ Å. Each of these eigenvectors has $3N$ components. We then evaluate the projection of every pair of these eigenvectors obtained at different $r_c$ values by the dot product, $\mathbf{p}_i(r_{c1}) \cdot \mathbf{p}_i(r_{c2})$. For eigenvectors that are slightly perturbed, this value is close to 1 and for orthogonal eigenvectors it is zero.

In figure 3, we exemplify the modifications in the lowest two non-trivial modes (i.e. modes 7 and 8) for the A chain of succinyl-CoA synthase (PDB code 1scu) by contour maps corresponding to these calculations. In this figure, the lower diagonal represents the projection of the eigenvectors of the slowest mode, and the upper diagonal is that for the second slowest mode. As is evident, both of these modes preserve their directionality for all $r_c > 11$ Å. For example, the eigenvectors representing the two slowest modes are identical for the $H_*$ matrix formed at $r_{c1} = 12$ Å, and its perturbed form obtained at $r_{c2} = 25$ Å (these two points are marked with a star sign on the figure).

The two most collective modes are found to yield similar results to those in Figure 3 for all proteins studied. However, it has been customary to follow the behavior of up to the first 20 slowest modes in protein calculations. It is therefore of interest to see the extent to which the eigenvectors, whose shapes are related to protein function in many studies, are sensitive to the choice of network construction. In Table 1, we report our calculations based on a set of 25 non-homolog proteins of various sizes. We find that, out of the 20 eigenvectors corresponding to the slowest eigenvalues obtained at $r_{c1} = 12$ Å, from 8 – 20 eigenvectors are approximately invariant to the choice of larger $r_{c2}$, exemplified here by $r_{c2} = 15$ Å (case A) and $r_{c2} = 18$ Å (case B). Our criterion for invariance is that the dot product $\mathbf{p}_i(r_{c1}) \cdot \mathbf{p}_i(r_{c2}) \geq 0.7$. Thus, in all the cases we examine, the two slowest non-trivial modes of motion are approximately invariant to matrix construction strategy. In contrast, the invariance of the remaining modes is highly dependent on the particular protein structure. There is a statistically significant dependence of the number of invariant modes on protein size (Pearson correlation is 0.57 and 0.58 for the respective cases). This reflects the fact that the increased number of redundant interactions contribute to the conditions outlined in the Appendix, leading to the separability of the essential and redundant parts of the Hessian, despite the increase in the number of surface residues. Note that the bond orientational parameter $\mathbf{Q}_6$ is essentially the same for the core and surface residues (data not shown).

**DISCUSSION**

In recent years, network models of proteins, RNA and their complexes have opened up previously unexplored areas of study, since the level of coarse graining adopted has been shown to describe several important phenomena unique to these self-assembled systems. The findings are mainly based on the observation that a simplified harmonic potential (equation 4) is capable of describing the collective modes of motion [7], which also are associated with the basic functioning of these molecular machines [35]. First, it was demonstrated that the Debye-Waller factors obtained from X-ray crystallography correlate with the fluctuations predicted by the theory [11]. This led to the study of the cross-correlations between the different parts of the system with confidence, leading to information not directly accessible by experiments; in particular, the coupled motions in the low frequency regions were found to shed light on many experimental findings and were utilized to uncover some mechanistic features; see, e.g. [36]. It was later shown that the eigenvectors associated with the lowest frequencies of motion also described the conformational changes accompanying binding [16,35,37,38,39,40]. The level of success in the latter work depends on the degree of collectivity displayed by the particular protein [22]. The number of modes that describe the essential motions is highly specific to the protein, or even to the different ligand bound forms of the same protein [21]. In yet other studies, the mode that best describes the conformational change was monitored to see if mutations of certain residues (carried out by modifying the force constant of contacting pairs, $\gamma_i$ in equation 4) affected the dominance of that particular mode [41,42]. Further, monitoring the response of the protein to local structural deformations [43] leads to valuable information on function and allosteric response [16,44].

The level of success of these studies, which all depend on the quality of the constructed Hessian, in relation to the method of network construction has not been addressed systematically. One exception is the work by da Silveira et al, which focuses on the residue contact properties for different network construction strategies [45]. Our analysis in the previous section uncovers spectral properties of the Hessian; in particular, the robustness of the most collective modes is based on properly including the local structure of proteins in the Hessian, whereas the longer range interactions build a redundant set. We discuss the implications of these findings by specific examples.

**Mean-square fluctuations of residues.** The residue-by-residue mean square fluctuations in a given protein are frequently exploited as a first step while constructing residue networks. The predictions of equation 6 are compared with the values obtained either experimentally or from MD simulations, and the $r_c$ value that best-represents the fluctuation profiles are selected to further study the system properties. We find for a set of 50 proteins that the correlation between the mean-square fluctuations of $C_\alpha$ atoms and the theoretical predictions of equation 6 improve as the cut-off distance is increased. This curious observation is valid up to $r_c$ values of at least 25 Å.

Two examples are displayed in figure 4: In figure 4a, we compare the mean square fluctuations calculated via MD simulations at 300 K of hen egg white lysozyme (HEWL, PDB code 6lyz, 129 residues) with the predictions from selected theoretical models of $r_c$ = 8, 16 and 25 Å. The details of the MD simulations are given in reference [46]. A lower correlation coefficient of 0.52 between the predicted and MD-calculated values is obtained at $r_c$ = 8 Å. This is due to the large fluctuations of residues 47-48 and 68-70 belonging to the loops within the β sheet region defined by residues (43-45, 51-53, 58-59, 64-65, 78-79). The important interactions of these loops with the rest of the protein are omitted at low $r_c$ values, leading to much larger fluctuation patterns than is actually present. At the higher $r_c$ values, exemplified by $r_c$ = 16 Å in figure 4a, these local interactions are restored, and the correlations of 0.90 is obtained. At $r_c$ > 18 Å the correlation is somewhat lowered due mainly to the suppressed fluctuations of the loop containing residues 47 and 48 when they are connected to long range residues (e.g. at $r_c$ = 25 Å in figure 4a, it is 0.76). Nevertheless, the mean square fluctuation profiles are correctly captured at all $r_c$ > 10 Å, with correlations of 0.75 or better. Also note that, of the slowest modes which have the largest contributions to the calculation of the fluctuations (equation 6), 10 have invariant eigenvectors for $r_c$ up to 18 Å for 6lyz (Table 1).

In figure 4b, we present another example for a 263 residue *β*-class protein (PDB code: 1arb), where the residue-by-residue experimental B-factors (bottom curve in gray) are compared with selected theoretical models: A relatively low correlation is obtained at $r_c$ = 8 Å; in particular, the fluctuations of surface loop residues 15 – 20 and 135 – 145 are overestimated due to the absence of important core-region contacts that are not taken into account at this cut-off distance. The $r_c$ = 16 Å model captures the experimentally determined fluctuation patterns, which remains unaltered at higher cut-offs. The Pearson correlation coefficients are 0.57, 0.81 and 0.82, at the respective cases displayed in figure 4b.

We emphasize that the behavior exemplified by figure 4 is not unique to these two proteins, but is rather a common property of all proteins. Thus, one may conveniently partition the Hessian into two (equation 8), where $H_*$ contains information due to the essential contacts of the matrix whereas $H_r$ is the residual part where the interactions are added in a symmetrical manner around the nodes beyond the adopted $r_c$ value (see the orientational order $Q_6$ in figure 1). The lowest eigenvalues are modified in a small window based on this partitioning, and the corresponding eigenvectors remain unchanged (figure 3 and Table 1). As a result, the inverse of the Hessian is nominally modified, accompanied by slight changes in the predicted $C_\alpha$ fluctuations (equation 6).

**A case study on predicting conformational change upon ligand binding.** Inasmuch as the global motions are dominated by one or the superposition of a few collective modes, the results for properties based-on these motions will not change appreciably with the different choices of $r_c$. Due to the invariance of the eigenvectors under a perturbation to the essential part of the Hessian, the mode based predictions on the direction of motion between the unbound and bound conformations of the protein are also expected to converge. An example is shown in figure 5 for the protein adenylate kinase (ADK), for which the eigenvector that belongs to the lowest eigenvalue is known to describe the conformational change with high accuracy due to the collective behavior of the hinge motion between its NMP binding (residues 30-67) and LID domains (residues118-167) [47].

The eigenvectors corresponding to the two slowest modes of motion of apo ADK are plotted on the PDB structure in figure 5a. The eigenvectors obtained at $r_c$ = 8 Å (orange) and $r_c$ =18 Å (green) are shown for both modes. The most collective mode is that of hinge bending of the LID (top) and NMP-binding (bottom) domains, and the next one is a twisting mode of these two domains. These modes are well separated at all $r_c$ values. In fact, mode 7 suffices to describe the conformational change upon ligand binding as shown in figure 5b. Here, we display in gray the residue-by-residue displacements of the apo and Ap5-bound structures of ADK obtained from the difference of their superposed experimental x-ray structures (PDB codes 4ake and 1ake, respectively). Also shown on the figure are the magnitudes of the eigenvector components acting on the residues from mode 7.

The Pearson correlation between the experimental and theoretical curves is 0.9 for all cut-off distances at and above $r_c = 8$ Å. The largest discrepancy between theory and experiment is observed in the region spanning NMP binding domain residues; the conformational change of the LID domain on the opposite side is faithfully reproduced irrespective of the choice of $r_c$. Moreover, the prediction of the overall conformational change does not change with $r_c$, although, it may be improved by the inclusion of additional modes.

**Shifts in eigenvalue ordering and its relation to physical interpretations.** The increase in the correlation coefficient with $r_c$ as well as its persistence to very high $r_c$ implies that the main ingredients that contribute to the fluctuation predictions are present in the Hessian obtained at a relatively low $r_c$, and the additional contacts act as a perturbation to this "essential" part of the matrix. However, one has to take extreme care in interpreting dynamical properties of proteins based-on the collective eigenvectors, as the ordering of global modes may shift with the selected cut-off distance as we exemplify next.

We begin by noting that, when matrices are perturbed by the addition of a diagonal matrix (as is approximated by the current systems and outlined in the Appendix), the eigenvalues of the perturbed matrix are interlaced; i.e., $\lambda_i(\mathbf{H}_* + \mathbf{H}_r) \in [\lambda_i(\mathbf{H}_*), \lambda_{i-1}(\mathbf{H}_*)]$; $i=2{:}n$ [see Section 8.1 in reference [33]]. In some cases, the gap between consecutive eigenvalues may be small enough so that their ordering changes. However, the associated eigenvectors are robust to the change in the eigenvalue order in the region where $\lambda_i$ is small. In figure 6, we exemplify one case where such a swapping of eigenvalue ordering occurs between the two slowest eigenvalues of HEWL. We first examine in figure 6a, the projection of the eigenvectors of the two most collective modes at different $r_c$ in the order they appear; e.g. $\mathbf{p}_7(r_c^1) \cdot \mathbf{p}_7(r_c^2)$ and $\mathbf{p}_8(r_c^1) \cdot \mathbf{p}_8(r_c^2)$. As in figure 3, the lower diagonal represents the projection of the eigenvectors of the slowest mode, and the upper diagonal is that for the second slowest mode. We find that the eigenvectors of each mode are parallel to each other in the cut-off ranges $8 < r_c^1, r_c^2 < 16$ and $16 < r_c^1, r_c^2 < 23$ Å. However, in the range $8 < r_c^1 < 16$ and $16 < r_c^2 < 23$ Å, the eigenvectors are nearly orthogonal for both modes. Evidently, at $r_c = 16$ Å, there is a change in the projection profile of vectors $\mathbf{p}_7(r_c > 16$ Å$)$ on $\mathbf{p}_7(r_c < 16$ Å$)$. We next recalculate the projections, assuming that the eigenvalue ordering is swapped at 16 Å due to interlacing; i.e. we calculate the projection $\mathbf{p}_7(r_c^1) \cdot \mathbf{p}_8(r_c^2)$ for the latter range of cut-off values. The "corrected" projections are shown in figure 6b. In this case, the effect of the perturbation on the slowest eigenvectors of the base Hessian is negligible (the dot product is 0.75 or better in all cases). We track the corresponding eigenvalues as a function of $r_c$ in figure 6c, where the swapping is evident.

We plot an eigenvector on the protein structure in figure 6d to visualize one of the major modes of action, and its robustness towards the choice of $r_c$. It corresponds to that labeled mode $i$ in figures 6b and 6c obtained at $r_c = 12$ and 18 Å (orange and green arrows, respectively). HEWL displays complex motions during its dynamics around the equilibrium state. Both of the major modes contributing to these motions act on the regions marked in red on the figure. They span the C-terminus residues and two loops joining β strands (43-45, 51-53) and (64-65, 78-79). The β sheet region is marked in yellow. The motion shown is a twisting of the loops; the motion not displayed also corresponds to the twisting of these loops, albeit in an orthogonal direction. Mode swapping and lack of dominance of a single mode exemplified here for HEWL suggest that, the complex motions in certain proteins may only be understood by studying the superposition of several such motions.

**Is there a recipe for residue network construction in protein models?** In this study, the degree of success of network models constructed from the PDB coordinates of proteins is shown to converge if the cut-off distance used is larger than a threshold value. We find that this value incorporates all the local essential interactions. A choice of $r_c$ in the vicinity of 16 Å covers the neighborhood structure of an arbitrary protein and its eigenvalue spectra [13]. However, for large proteins, this will introduce a large number of interactions which will be computationally demanding during the matrix inversion procedure. In such cases, one may resort to compute bond orientational order parameters and choose the optimal value when the structural descriptor $Q_6$ converges. For large proteins the number of nodes will be high enough to obtain statistics for smooth curves where the peaks may be discerned, a problem that cannot be circumvented for small system sizes. However, this approach is computationally demanding for large systems and higher cut-off values; e.g. to calculate $Q_6$ in the $r_c$ range of 5 – 28 Å on a computer with 2.5 GHz Intel Quad core CPU and 8GB RAM for the protein PDB codes 6lyz, 1ad2 and 1scu takes 740, 1700, and 3000 seconds, respectively. The matrix Frobenius norm of the scaled Hessian, especially for large systems constitutes an attractive alternative route due to its low computational cost and straightforward implementation.

Thus, for any given protein, monitoring any of these three measurables, i.e. the eigenvalue spectra, orientational order, or Frobenius norm, will give an idea on the range of values that may be used for $r_c$. Nevertheless, it is advisable to track the inner products of eigenvectors to ascertain that the modes of interest are robust to this selection. Furthermore, if the properties of interest rely on the predictions based on a few selected modes, one must also track the eigenvectors in case interlacing of the eigenvalues occurs. A quick scan of the projections of a given mode obtained at different network constructions will reveal such anomalies.

In summary, we have shown that the slow modes are robust to the details of network construction once the essential contacts in the first few coordination shells are included. Therefore, the properties that depend on the most collective modes may be studied independent of this choice. This is in contrast to the modes that affect the medium to high frequency motions, since interlacing and mode shifts will lead to unpredictable changes in the eigenvectors. Therefore, in studies deriving information by relying on the superposition of a large number of modes, developing a sound network construction strategy is essential.

**APPENDIX**

We have previously shown [12,13] that the Hessian may be decomposed into the product,

$$\mathbf{H} = \mathbf{B}\mathbf{B}^\mathrm{T} \qquad (A1)$$

where **B** is the $3N \times M$ direction cosine matrix. The overall interactions are also written as the sum of the $\mathbf{B}_*$ and $\mathbf{B}_r$ matrices, containing the essential and the residual interactions, respectively. By substitution into equation 8, the Hessian may thus be expressed as

$$\begin{aligned}\mathbf{H} &= (\mathbf{B}_* + \mathbf{B}_r)(\mathbf{B}_* + \mathbf{B}_r)^\mathrm{T} \\ &= \mathbf{B}_*\mathbf{B}_*^\mathrm{T} + \mathbf{B}_*\mathbf{B}_r^\mathrm{T} + \mathbf{B}_r\mathbf{B}_*^\mathrm{T} + \mathbf{B}_r\mathbf{B}_r^\mathrm{T}\end{aligned} \qquad (A2)$$

H may also be viewed as an $N \times N$ supermatrix (equation 4) whose *ij*th element is the 3×3 matrix $\mathrm{H}^{ij} = \mathrm{H}_*^{ij} + \mathrm{H}_r^{ij}$.

The cross terms in equation A2, $(\mathbf{B}_*\mathbf{B}_r^T)$ and $(\mathbf{B}_r\mathbf{B}_*^T)$ are each zero. To see why, consider the $\mathbf{B}_*$ matrix with dimension $3N \times M$ whose last ($r$) columns have zero elements and the $\mathbf{B}_r^T$ matrix with dimension $M \times 3N$ whose first ($M-r$) rows have zero elements. Then,

$$\mathbf{B}_*\mathbf{B}_r^T = \begin{bmatrix} \cdot & \cdot & \cdot & | & 0 & 0 \\ \cdot & \cdot & \cdot & | & \vdots & \vdots \\ \cdot & \cdot & \cdot & | & 0 & 0 \end{bmatrix}_{3N \times M} \begin{bmatrix} 0 & 0 & 0 \\ \vdots & \vdots & \vdots \\ 0 & 0 & 0 \\ - & - & - \\ \cdot & \cdot & \cdot \\ \cdot & \cdot & \cdot \end{bmatrix}_{M \times 3N} = \mathbf{0} \quad (A3)$$

Here, if bead $i$ participates in the $m^{th}$ interaction, then the terms $(3i,m)$, $(3i+1,m)$, and $(3i+2,m)$ are non-zero and consist of the direction cosine of bead $i$ along interaction $m$, in the $x$-, $y$-, and $z$-directions, respectively. They are zero otherwise. A similar argument holds for $(\mathbf{B}_r\mathbf{B}_*^T)$.

We now denote the elements of the $\mathbf{B}_*$ and $\mathbf{B}_r$ matrices by $\cos\alpha_{ij}^p$ and $\cos\beta_{ij}^p$, respectively, where $p$ is the $x$-, $y$-, or the $z$-direction. In addition, the direction cosines for the interaction between beads $i$ and $j$ have the relationship $\cos\alpha_{ij}^p = -\cos\alpha_{ji}^p$. Then, the elements of the remaining terms in equation A2 may be explicitly written. For the essential component we have the terms,

$$\mathbf{H}_*^{ij} = (\mathbf{B}_*\mathbf{B}_*^T)^{ij} = \begin{bmatrix} -\cos^2\alpha_{ij}^x & -\cos\alpha_{ij}^x\cos\alpha_{ij}^y & -\cos\alpha_{ij}^x\cos\alpha_{ij}^z \\ -\cos\alpha_{ij}^x\cos\alpha_{ij}^y & -\cos^2\alpha_{ij}^y & -\cos\alpha_{ij}^y\cos\alpha_{ij}^z \\ -\cos\alpha_{ij}^x\cos\alpha_{ij}^z & -\cos\alpha_{ij}^y\cos\alpha_{ij}^z & -\cos^2\alpha_{ij}^z \end{bmatrix} \quad (A4)$$

$$\mathbf{H}_*^{ii} = (\mathbf{B}_*\mathbf{B}_*^T)^{ii} = \begin{bmatrix} \sum_j\cos^2\alpha_{ij}^x & \sum_j\cos\alpha_{ij}^x\cos\alpha_{ij}^y & \sum_j\cos\alpha_{ij}^x\cos\alpha_{ij}^z \\ \sum_j\cos\alpha_{ij}^x\cos\alpha_{ij}^y & \sum_j\cos^2\alpha_{ij}^y & \sum_j\cos\alpha_{ij}^y\cos\alpha_{ij}^z \\ \sum_j\cos\alpha_{ij}^x\cos\alpha_{ij}^z & \sum_j\cos\alpha_{ij}^y\cos\alpha_{ij}^z & \sum_j\cos^2\alpha_{ij}^z \end{bmatrix} \quad (A5)$$

For the residual component, we have similar matrix elements for $\mathbf{H}_r^{ij}$ and $\mathbf{H}_r^{ii}$ with $\cos\beta_{ij}^p$ replacing $\cos\alpha_{ij}^p$ terms. We now assume that the elements of the $\mathbf{H}_*$ matrix are dependent on each other due to the underlying local structure of the protein, whereas those of the $\mathbf{H}_r$ matrix are independent. For a sufficiently large number ($r$) of residual interactions appearing in such independent directions, we have the following approximate values for the diagonal super-elements of the residual component:

$$(\mathbf{B}_r\mathbf{B}_r^T)^{ij} = \begin{cases} \sum_j\cos^2\beta_{ij}^p \approx r_i/3 & i=j; p=q \\ \sum_j\cos\beta_{ij}^p\cos\beta_{ij}^q \approx 0 & i=j; p \neq q \\ -\cos^2\beta_{ij}^p = [-1,0] & i \neq j; p=q \\ -\cos\beta_{ij}^p\cos\beta_{ij}^q = [-1,1] & i \neq j; p \neq q \end{cases} \quad (A6)$$

with $r_i$ being the number of redundant interactions of each residue and $\sum_i r_i = 2r$. In addition, there is the constraint due to the law of cosines, $\cos^2\beta_{ij}^x + \cos^2\beta_{ij}^y + \cos^2\beta_{ij}^z = 1$. Substituting these limits, the off-diagonal and diagonal elements of the super-matrix elements of $\mathbf{H}_r$ are in the bounds

$$\mathbf{H}_r^{ij} = (\mathbf{B}_r\mathbf{B}_r^T)^{ij} \approx \begin{bmatrix} [-1,1] & [-1,0] & [-1,0] \\ [-1,0] & [-1,1] & [-1,0] \\ [-1,0] & [-1,0] & [-1,1] \end{bmatrix} \quad (A7)$$

$$\mathbf{H}_r^{ii} = (\mathbf{B}_r \mathbf{B}_r^{\mathrm{T}})^{ii} \approx \begin{bmatrix} r_i/3 & 0 & 0 \\ 0 & r_i/3 & 0 \\ 0 & 0 & r_i/3 \end{bmatrix} \tag{A8}$$

For the cut-off distances considered in this study, $r_i > 10$. For example, for the protein with PDB code 1sca which has 287 nodes, if the essential part of the Hessian is constructed at $r_c = 12$ Å, and the perturbed Hessian is constructed at 16 Å, the average number of residual interactions per node is 33. Thus, the leading terms of $\mathbf{H}_r$ are along the diagonal and it may be approximated by a diagonal matrix. The effect of the perturbations brought about by the $\mathbf{H}_r$ matrix with such approximate bounds, on the eigenvalues and the corresponding eigenvectors of the $\mathbf{H}_*$ matrix are thoroughly investigated in the text with accompanied examples.

**ACKNOWLEDGEMENTS.** We thank Ibrahim Inanc for useful discussions.

**FIGURE LEGENDS**

**Figure 1.** Bond orientational order parameter, $Q_6$, computed in the cut-off distance range $r_c = 5 - 28$ Å with 1 Å increments for three example proteins. The sizes and PDB codes are indicated on the figures; $Q_6$ averaged over 30 non-homolog proteins of varying sizes is also displayed (gray line). We observe a decrease from $Q_6 = 0.5$ at low $r_c$ towards a convergence to $Q_6 = 0.1 - 0.2$ as $r_c$ is increased. $Q_6$ is a geometrical analysis which follows from the local force balance conditions due to the anisotropy of contacts in ANM.

**Figure 2.** The matrix Frobenius norm of the scaled Hessians (equation 11) computed for six example proteins, scanned at $r_c$ values with 1 Å increments. The sizes and PDB codes are indicated on the figures. The minima are marked by the arrows.

**Figure 3.** Inner product maps for the eigenvectors $\mathbf{p}_7$ and $\mathbf{p}_8$ corresponding to the slowest two modes of the protein with PDB code 1scu. These are labeled modes 7 and 8, since the first six modes correspond to translational and rotational motions. At each point we compute the inner product of the eigenvectors of mode $i$ at a pair of $r_c$ values; i.e. $\mathbf{p}_i(r_{c1}) \cdot \mathbf{p}_i(r_{c2})$. Lower triangle is for mode 7 and upper one is for mode 8. We find that above $r_c = 9$ Å global modes remain unaltered as more contacts are added with larger cut-off radii. For example, the value of the projection $\mathbf{p}_7$ of the computed from networks formed at $r_c = 12$ and 22 Å is 0.99 and that for $\mathbf{p}_8$ is 0.96 (these points are marked by star signs on the figure).

**Figure 4.** Fluctuation profiles ($\Delta\mathbf{R}_i \cdot \Delta\mathbf{R}_i$); calculated by equation 6 and using three different residue network models obtained at $r_c = 8, 16, 25$ Å are compared with; **(a)** residue fluctuations of HEWL calculated from MD simulations (lower gray curve). **(b)** X-ray B-factors (lower gray curve) of the 268 residue achromobacter lyticus protease (PDB code: 1arb).

**Figure 5. (a)** The two most collective modes of apo ADK mapped onto the structure. For this protein, eigenvectors obtained at $r_c = 8$ and 18 Å (orange and green arrows, respectively) are nearly indistinguishable. Mode 7, which describes hinge bending, is dominant and accounts for most of the conformational change upon binding as shown in **(b)** where the displacement profiles between the unbound and bound forms (PDB codes: 4ake and 1ake, respectively) are explored. The experimental displacements are shown in gray. Predictions from the relative magnitudes of the eigenvector corresponding to the slowest mode obtained at $r_c = 8, 10, 15$ Å are shown in black. The latter curves are displaced to guide the eye, and their zero baselines are marked by the dotted curves. The Pearson correlation between the experimental curve and each of the predictions is 0.9.

**Figure 6.** Inner product maps $\mathbf{p}_i(r_c^1) \cdot \mathbf{p}_i(r_c^2)$ for eigenvectors $\mathbf{p}_7$ and $\mathbf{p}_8$ of HEWL displayed in **(a)** show that there is possible swapping in the mode order at $r_c = 16$ Å; this prediction is justified in **(b)** where the dot product $\mathbf{p}_i(r_c^1) \cdot \mathbf{p}_j(r_c^2)$ for the regions that are swapped restores the picture with the slow modes staying unaltered with the choice of $r_c$ beyond a lower threshold. The swapping of the modes may also be tracked in the eigenvalues as shown **(c)**. In **(d)**, the mode labeled $i$ in parts **b** and **c** obtained is mapped onto the protein structure for $r_c = 12$ and 18 Å (orange and green arrows, respectively). The mode is most effective in the regions colored red: The loops joining the yellow colored β-sheet structure and the C-terminus. Mode $j$ (not shown) also acts on the same regions, albeit in orthogonal directions.

**Table 1.** Number of approximately invariant eigenvectors[1,2] in matrices $H_*$ and $H_*+H_r$.

| PDB id | no. of residues | # of eigenvectors | |
|---|---|---|---|
| | | Case A[3,4] | Case B[3,5] |
| 5rxn | 54 | 13 | 11 |
| 1lfb | 77 | 15 | 12 |
| 1tpg | 91 | 17 | 10 |
| 1fkj | 107 | 11 | 9 |
| 2pii | 112 | 12 | 11 |
| 1hce | 118 | 8 | 8 |
| 1bp2 | 123 | 12 | 10 |
| 6lyz | 129 | 11 | 10 |
| 1ash | 147 | 13 | 9 |
| 1rfj | 148 | 18 | 17 |
| 1ilk | 151 | 18 | 11 |
| 1hbq | 176 | 9 | 10 |
| 1cus | 197 | 14 | 11 |
| 1gen | 200 | 10 | 9 |
| 1ad2 | 224 | 15 | 12 |
| 1fib | 249 | 11 | 10 |
| 1dih | 272 | 18 | 14 |
| 1scu | 287 | 19 | 13 |
| 1ghr | 306 | 15 | 12 |
| 1ede | 310 | 13 | 8 |
| 1cem | 363 | 14 | 10 |
| 1kaz | 378 | 17 | 12 |
| 1inp | 400 | 16 | 15 |
| 2bnh | 456 | 17 | 17 |
| 1vnc | 576 | 15 | 12 |
| 1sly | 618 | 20 | 20 |
| 1aa6 | 696 | 19 | -[6] |
| 1kit | 757 | 19 | -[6] |

[1] Excluding the first six eigenvectors corresponding to translation and rotation of the whole molecule
[2] We report the number of eigenvectors with the lowest 20 eigenvalues in the $H_*$ matrix whose projection on $H_*+H_r$ is larger than the set threshold of 0.7.
[3] $H_*$ is formed at $r_{c1} = 12$ Å
[4] $H_*+H_r$ is formed $r_{c2} = 15$ Å
[5] $H_*+H_r$ is formed $r_{c2} = 18$ Å
[6] Not computed

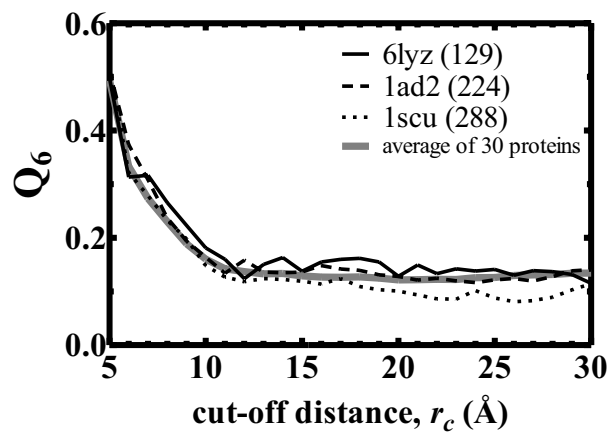

*Figure 1*

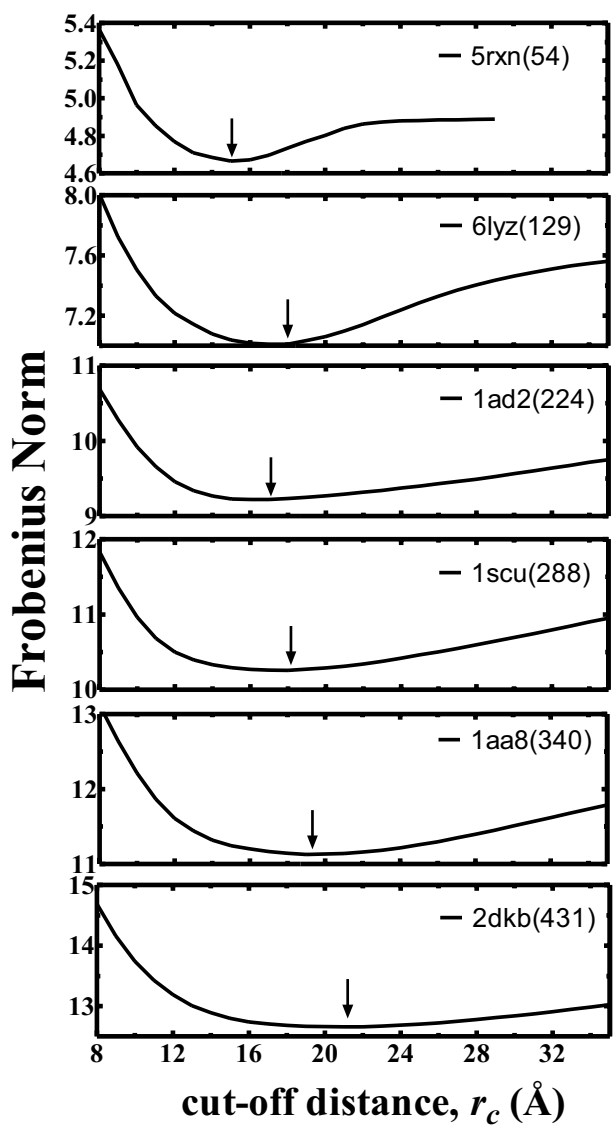

*Figure 2*

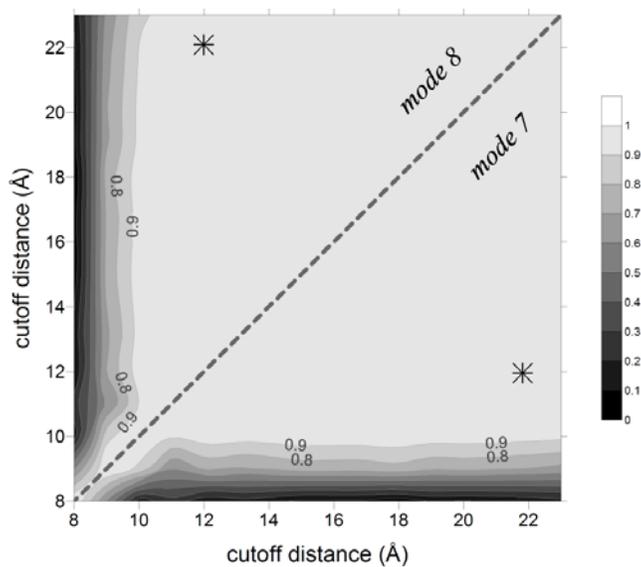

*Figure 3*

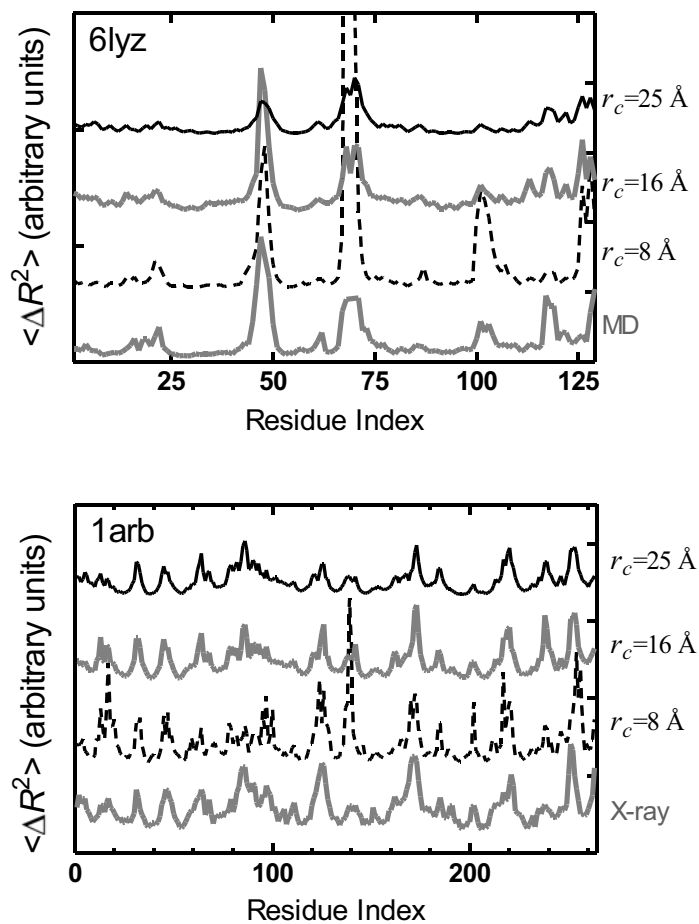

*Figure 4*

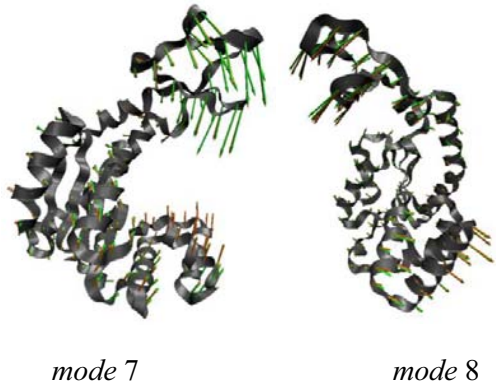

*mode* 7            *mode* 8

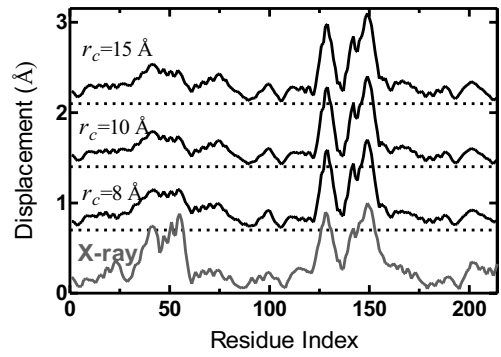

*Figure 5*

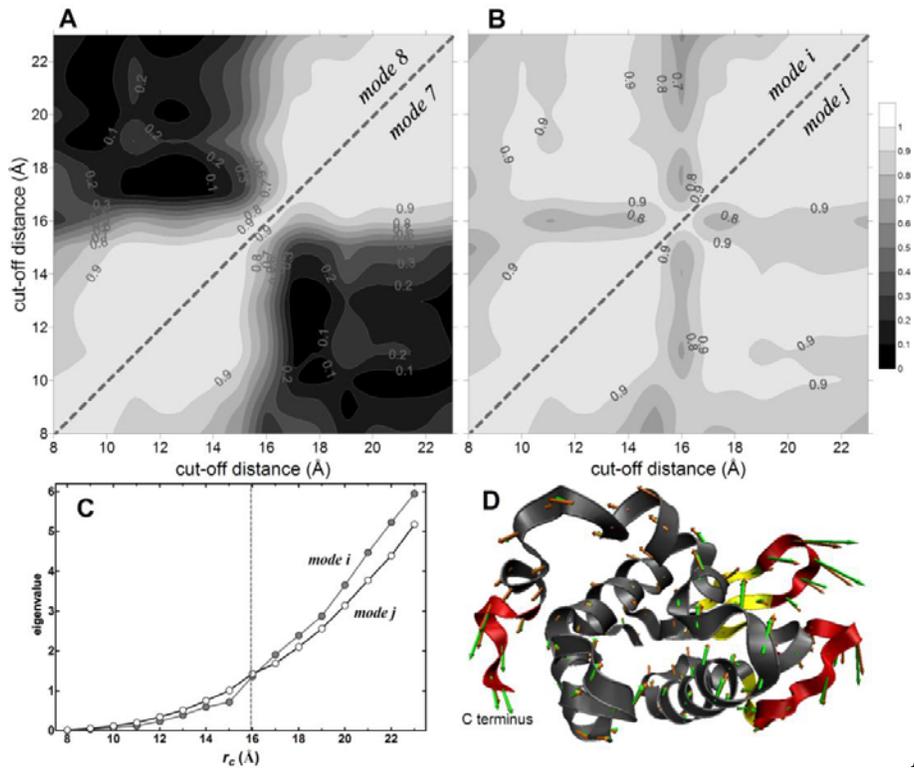

*Figure 6*